\documentclass[%
 reprint,
 superscriptaddress,
 amsmath,amssymb,
 aps,
]{revtex4-2}

\usepackage{graphicx}
\usepackage{dcolumn}
\usepackage{bm}
\usepackage{hyperref}
\hypersetup{
  colorlinks   = true,
  urlcolor     = blue, 
  linkcolor    = black, 
  citecolor   = blue
}
\usepackage[all]{hypcap}
\usepackage{color}

\begin{document}

\preprint{APS/123-QED}

\title{Anisotropic hybridization in CeRhSn\\}

\author{Thomas U. B\"ohm}
\affiliation{Los Alamos National Laboratory, Los Alamos, New Mexico 87545, USA}

\author{Nicholas S. Sirica}
\affiliation{Los Alamos National Laboratory, Los Alamos, New Mexico 87545, USA}
\affiliation{U.S. Naval Research Laboratory, Washington, DC 20375, USA}

\author{Bo Gyu Jang}
\affiliation{Los Alamos National Laboratory, Los Alamos, New Mexico 87545, USA}
\affiliation{Department of Advanced Materials Engineering for Information \& Electronics, Kyung Hee University, Yongin 17104, Republic of Korea}

\author{Yu Liu}
\affiliation{Los Alamos National Laboratory, Los Alamos, New Mexico 87545, USA}

\author{Eric D. Bauer}
\affiliation{Los Alamos National Laboratory, Los Alamos, New Mexico 87545, USA}

\author{Yue Huang}
\affiliation{Los Alamos National Laboratory, Los Alamos, New Mexico 87545, USA}

\author{Christopher C. Homes}
\affiliation{National Synchrotron Light Source II, Brookhaven National Laboratory, Upton, New York 11973, USA}

\author{Jian-Xin Zhu}
\affiliation{Los Alamos National Laboratory, Los Alamos, New Mexico 87545, USA}

\author{Filip Ronning}
\affiliation{Los Alamos National Laboratory, Los Alamos, New Mexico 87545, USA}

\date{\today}
\begin{abstract}
The optical conductivity $\sigma(\omega,T)$ of CeRhSn was studied by broadband infrared spectroscopy. Temperature-dependent spectral weight transfer occurs over high energy ($0.8\,$eV) and temperature (${\sim}500\,$K) scales, classifying CeRhSn as a mixed valent compound. The optical conductivity reveals a substantial anisotropy in the electronic structure. Renormalization of $\sigma(\omega,T)$ occurs as a function of temperature to a coherent Kondo state with concomitant effective mass generation. Associated spectroscopic signatures were reproduced remarkably well by the combination of density functional theory and dynamical mean field theory using a momentum-independent self energy. The theory shows that the anisotropy for energies $>10\,$meV is mainly driven by the bare three-dimensional electronic structure that is renormalized by local electronic correlations. The possible influence of magnetic frustration and quantum criticality is restricted to lower energies. 
\end{abstract}

\maketitle
Quantum phase transitions in lanthanide and actinide based heavy fermions are some of the most widely studied phenomena in condensed matter physics \cite{pottgen2015, janka2016, pottgen2016, kirchner2020, lohneysen2007, si2010}. Here, varying the hybridization strength between localized $f$- and delocalized conduction electrons (c) allows for the continuous tuning between strongly $f$-c hybridized Kondo and weakly hybridized magnetic states through use of nonthermal control parameters such as pressure, chemical composition, or magnetic field. Within this canonical description of quantum criticality for $f$-block materials \cite{doniach1977}, mixed-valent compounds are frequently overlooked as the energy scale for valence fluctuations dominates over the low energy competition between magnetic order and Kondo screening, leading to the emergence of many-body Kondo singlet states at a characteristically high temperature ($>100\,$K) \cite{shimura2021, cornelius2002, kim2003}. However, the introduction of geometric frustration and/or reduced dimensionality between Kondo ions adds another dimension to the global magnetic phase diagram of $f$-electron materials \cite{coleman2010, si2006, zhao2019}, providing an alternate route to realize quantum criticality, even in the mixed-valent compounds \cite{matsumoto2011, grbic2022, tokiwa2015}. 

\begin{figure}
\includegraphics{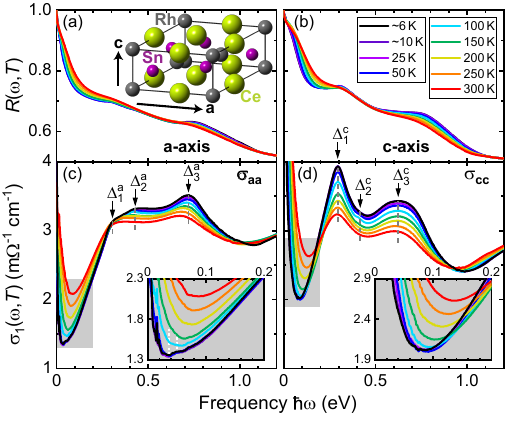}
\caption{Experimental results for CeRhSn. The reflectivity $R(\omega,T)$ in (a) and (b) and real optical conductivity $\sigma_1(\omega,T)$ in (c) and (d) are plotted as a function of frequency and temperature. Panels (a) and (c) compile the spectra for a polarization applied along the $a$-axis, panels (b) and (d) for a polarization along $c$. The insets of panel (c) and (d) enlarge the spectra inside the grey-shaded boxes. Dashed white lines indicate the minimum $\omega_\mathrm{min}$ and shoulder $\omega_\mathrm{s}$ developing at low temperatures. Narrow features below the minimum are optical phonon modes. The color-coding for the temperatures is preserved for all experimental data throughout the manuscript. The Supplemental Material provides data in the full frequency range up to $2.7\,$eV \cite{supp}.}
\label{experiment}
\end{figure}

The CeTX  (T=transition metal, X=p-block element) family of intermetallics is a promising material class for investigating such an interplay between geometric frustration and Kondo screening in a mixed valent-compound \cite{pottgen2015}. Here, Ce atoms sit on the distorted kagome lattice of a hexagonal ZrNiAl-type ($P\bar{6}2m$) structure depicted in Fig.~\ref{experiment}(a). Of the members in this family, CeRhSn has attracted significant attention due to the rich variety of physical properties exhibited by this compound, including mixed valency \cite{slebarski2002, sundermann2021, gamza2009}, geometric frustration, quantum criticality \cite{tokiwa2015, kittaka2021}, and strong magnetic and electronic anisotropy \cite{kim2003}. Ising-like susceptibilities $\chi_c \approx 10\chi_a$ at low temperatures naively contradict the mixed valent behavior, which would lead to near equally populated crystal field levels. Meanwhile, resistivity measurements reveal signatures of Kondo coherence in the $ab$-plane that are notably absent along the crystallographic $c$-axis, suggesting an anisotropic mass renormalization in momentum space. Such behavior is at odds with the majority of mixed-valent compounds \cite{bucher1996, okamura2004}, whose isotropic, cubic symmetry often support a Fermi-liquid ground state, and emphasizes the impact that reduced dimensionality and magentic frustration may have on Kondo coherence. 

We perform Fourier transform infrared (FTIR) spectroscopy complemented by density functional theory (DFT) plus dynamical mean-field theory (DMFT) \cite{jang2023, haule2007, haule2010}. Experimentally, a polarization-dependent measurement separates the $a$ and the $c$-axis optical conductivity $\sigma_{aa}$ and $\sigma_{cc}$. Emergent spectroscopic features indicate a transition from a high temperature state with fluctuating local moments to a low temperature coherent anisotropic Kondo state,  governed by the screening of $f$-electron moments by conduction electrons. Their hybridization leads to an energy gap which introduces a multitude of new inter-band absorption channels at infrared energy scales, directly observable by our experiment \cite{kimura2021}. It also yields flat bands populated with heavy fermions, where intra-band transitions appear as a Drude peak and provide information on the mass enhancement and scattering rate~\cite{puchkov1996}. The three-dimensional multi-band electronic structure, however, complicates a conclusive analysis from the experiment alone. Therefore, we applied DFT+DMFT to resolve the spectroscopically observed features. Based on a momentum-independent self-energy from localized $f$-states, we gain insight into the mechanism of hybridization and effective mass enhancement. We find that the anisotropy in the optical conductivity can be reproduced by applying local Kondo physics to the framework of an intrinsically anisotropic electronic structure reflected in the DFT.

Experimentally, single crystal CeRhSn was grown via the Czochralski method \cite{czochralski1918} in a tri-arc furnace under high-purity argon atmosphere \cite{kim2003}. The orientation of the as-cast single crystalline rod was determined by means of Laue diffraction and verified within our experiments by using infrared active optical phonons as a spectroscopic indicator for lattice orientation (see Fig.~\ref{supplement1} in the Supplemental Material \cite{supp}). Such characterization demonstrates this sample to be single grain throughout the measured $ac$-plane. Reflectivity experiments were performed at near normal incidence ($< 15^{\circ}$) by applying linearly polarized light parallel to the crystallographic $a$ and $c$ axes \cite{homes1993}. We obtained the reflectivity spectra as displayed in Figs.~\ref{experiment}(a) and (b), respectively. The optical conductivity, $\sigma(\omega,T)$, was determined from the experimental reflectivity spectra by way of a Kramers-Kronig transform \cite{tanner2015, henke1993, gulliksonWebsite, supp}. This  
yields the real parts of the optical conductivity tensor elements $\sigma_{aa}(\omega,T)$ and $\sigma_{cc}(\omega,T)$ as presented in Figs.~\ref{experiment}(c) and (d), respectively. Details of the experiment and analysis are described in the Supplemental Material \cite{supp}. 

We observe that both directions of 
\begin{figure}[b]
\includegraphics{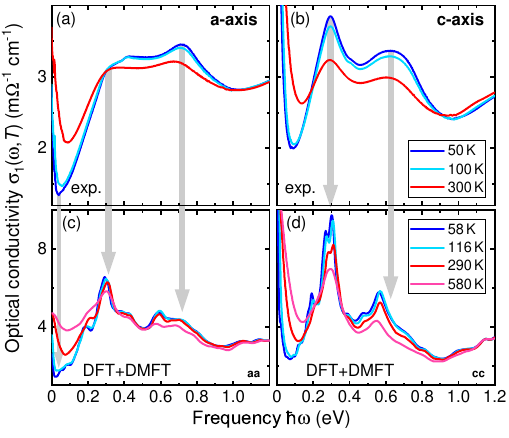}
\caption{Comparison of experimental optical conductivity with theoretically computed optical conductivity. Panels (a) and (c) duplicate the optical conductivity of Fig.~\ref{experiment} for a subset of temperatures, where nearby theoretical calculations, shown in (b) and (d) are available. Grey arrows mark significant experimentally determined peak-positions and the minimum in (a) for the lowest temperature.}
\label{theory}
\end{figure}
the optical conductivity $\sigma_{aa}$ and $\sigma_{cc}$ have a clear dip around $0.1\,$eV followed by three broad but distinct mid-infrared (MIR) peaks labeled $\Delta_i$ (as marked in Figs.~\ref{experiment}(c) and (d)) that extend from $0.2\,$eV up to $0.8\,$eV. The peak at $\Delta_2$ becomes visible below $150\,$K. The energies of $\Delta_1$, $\Delta_2$, and $\Delta_3$ are similar for both directions, while the intensities of these features are clearly anisotropic as could be expected for interband transitions in a 
\begin{figure*}
\includegraphics{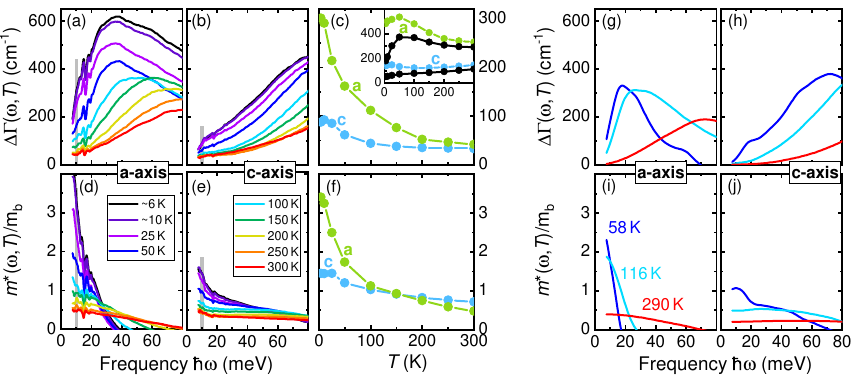}
\caption{Extended Drude analysis for the scattering rate and effective mass. Panels (a) and (b) show the relative scattering rates for the $a$ and $c$ axis, (d) and (e) the effective mass, respectively. Averaged values from $9\textnormal{-}11\,$meV are compiled as a function of temperature in (c) and (f). The inset of (c) depicts absolute values of the averaged scattering rate $\Gamma(9\textnormal{-}11\,\mathrm{meV},T)$ along with $\Gamma_0(T)$ (black) from Ref.~\cite{kim2003}. Panels (g-j) are the respective results, obtained from theoretically computed optical conductivity $\sigma_{aa}$ and $\sigma_{cc}$ shown in Figs.~\ref{theory}(c) and (d).}     
\label{extDrudeParameters}
\end{figure*}
hexagonal crystal. The most striking aspect of the data is a shift of spectral weight from low to high energies with decreasing temperature. In general, a shift of spectral weight is a clear signature of an energy gap $\Delta$ with optically-induced transitions of quasiparticles across the gap. The transfer of spectral weight to high energies is more prominent in $\sigma_{cc}$ than in $\sigma_{aa}$. The spectral weight transfer does not saturate at room temperature. Rather, an extrapolation of the continuing spectral weight loss gives a temperature scale of $500 \pm 100\,$K \cite{supp}, which we identify as the single ion Kondo scale. In contrast, low energy features are more striking in $\sigma_{aa}$ than in $\sigma_{cc}$. A new minimum at $\omega_\mathrm{min}=35\,$meV followed by a shoulder at $\omega_\mathrm{s}=50\,$meV develops at $100\,$K only for $\sigma_{aa}$, and is most pronounced and saturated for $T \leq 50\,$K (Fig.~\ref{experiment} insets). 

To elucidate the origin of the spectral weight transfer and observed anisotropies in the optical conductivity we have performed electronic structure calculations by fully charge self-consistent DMFT, combined with DFT~\cite{haule2010} using the WIEN2k code \cite{Blaha2020}. The Perdew-Burke-Ernzerhof generalized gradient approximation (PBE-GGA) was employed for the exchange-correlation potential~\cite{Perdew1996}. Electronic correlation effects of Ce 4$f$ orbitals were treated by a momentum-independent (local) self-energy from the DMFT part with a continuous time quantum Monte Carlo (CTQMC) solver~\cite{haule2007}. Hubbard parameters $U=5\,$eV, and $J=0.7\,$eV were used. The optical conductivity was computed with the formalism presented in Ref.~\cite{haule2010} (see Supplemental Material for details \cite{supp}). A comparison of the experimentally measured and theoretically calculated optical conductivity is presented in Fig.~\ref{theory}. Remarkable similarities are apparent. The previously described peak structure and spectral weight shift, which is more pronounced in $\sigma_{cc}$, is by and large captured by the theory. The more pronounced minimum in $\sigma_{aa}$ at low temperatures/energies is also captured by the theory. The good agreement implies that the observed temperature dependent anisotropies stem from a renormalization of the electronic structure that is local in real space, yielding a momentum independent self energy that is captured within the DMFT framework. 

We now turn our attention to the lowest energies accessible by our experiment to gain insight on the dynamics of the heavy fermion state. Analyzing the data using an extended Drude model enables the extraction of a frequency dependent scattering rate $\Gamma(\omega,T) =  \omega^2_\mathrm{p}/(4\pi)\,\mathrm{Re}[1/\sigma(\omega,T)]$ and effective mass $m^\ast(\omega,T)/m_\mathrm{b} = -\omega^2_\mathrm{p}/(4\pi\omega)\,\mathrm{Im}[1/\sigma(\omega,T)]$, with the bare band mass $m_\mathrm{b}$ \cite{puchkov1996}. The plasma frequency $\omega_\mathrm{p}$ is given by the carrier density through the optical conductivity sum rule. A precise value is difficult to determine due to the presence of interband transitions. We chose $\omega_\mathrm{p}=1\,$eV and note that a different value will only scale our results described below, as we limit our analysis to systematic trends that are independent of the precise value of the plasma frequency. Figs.~\ref{extDrudeParameters}(a-c) shows $\Delta\Gamma = \Gamma-\Gamma_0$ with the zero-frequency scattering rate $\Gamma_0(T) = \omega^2_\mathrm{p}/(4\pi)\,\rho_\mathrm{dc}(T)$. The difference allows for a better view of the frequency dependence beyond $\rho_\mathrm{dc}$. The results for the effective mass $m^\ast$ are shown in panels (d-f). The effective mass is strongly frequency and temperature dependent in the low energy limit ($\lesssim 40\,$meV), which is consistent with strong electronic renormalization observed in transport and thermodynamic measurements \cite{slebarski2002, kim2003}. We obtain increasingly larger scattering rates and effective masses in the $ab$-plane than along the $c$-axis with decreasing temperature. Comparable values at $300\,$K turn into a disproportional growth in the $ab$-plane particularly below ${\sim}150\,$K, reaching $\Delta\Gamma_a = 3.3\,\Delta\Gamma_c$ and $m^\ast_a = 2.4\,m^\ast_c$ at ${\sim}6\,$K. This anisotropy is consistent with the previous observation that the minimum in the optical conductivity is driven to lower energies for the $ab$-plane data relative to the $c$-axis data.

As mentioned in the introduction, CeRhSn presents a confounding confluence of mixed valent energy scales, strongly anisotropic low temperature transport and thermodynamic responses as well as quantum criticality, which raises the question as to what is the role of the reduced dimensionality and/or magnetic frustration present in the Ce-lattice of this compound. In a simple mean-field picture of the Kondo lattice, a hybridization gap forms between the conduction bands and a renormalized heavy band \cite{shim2007, kimura2021}. Precisely such a hybridization gap is observed in the optical conductivity response of many Ce-lattice compounds \cite{okamura2002, bucher1996, okamura2004, mena2005, kimura2011, kimura2011a, chen2016}. In CeRhSn the dominant MIR peak occurs for the $c$-axis polarization at $0.3\,$eV, consistent with a previous report \cite{okamura2008}. This energy scale is larger than other well-known mixed valent compounds such as YbAl$_3$, CeSn$_3$, and CePd$_3$ \cite{okamura2004, bucher1996}, consistent with the mixed valent classification for CeRhSn from other spectroscopic probes \cite{slebarski2002, gamza2009, sundermann2021}. Furthermore, it is clear that the transfer of spectral weight will continue well above room temperature. Examining the temperature dependence of the spectral weight contained in the peaks suggests this spectral weight transfer will persist up to $500\pm100\,$K \cite{supp}, giving a Kondo temperature similar to other mixed valent compounds. 

It is notable that the spectral weight transfer extends to energies as large as $0.8\,$eV, with a clear second subdominant peak near $0.6\,$eV, and a strongly anisotropic response between the $a$- and $c$-axes conductivity. At first glance this could suggest a distribution of energy scales in either real or momentum space. These spectral features, however, are all reproduced in our DFT+DMFT calculations shown in Fig.~\ref{theory}. Single site DMFT assumes a momentum independent self energy of the $f$-electrons. The resulting anisotropy, and distribution of energy scales is thus a consequence of the underlying band structure. Each band and $\bm k$-point will contribute differently to the directional dependent conductivity based on matrix elements for the individual optical transitions, and each of those $\bm k$-points/bands will have a different hybridization strength to the $f$-orbital, which is responsible for the electronic renormalization. Thus, the renormalization of CeRhSn’s electronic structure at high energies appears similar to other mixed valent compounds, but with the added complexity of anisotropic $f$-c hybridization present in the band structure. 

Concomitant with the formation of the hybridization gap, the low frequency conductance will reflect the scattering of the conduction electrons off of the $f$-moments. At high temperatures the scattering occurs off of individual $f$-moments, while below the coherence temperature, there is a collective screening of the local moments and an associated mass enhancement of the quasiparticles at the Fermi energy. It is generally difficult to resolve the coherence temperature in spectroscopic measurements. Previous optical conductivity measurements on YbAl$_3$ have associated the emergence of a peak within the hybridization gap as a signature of lattice coherence \cite{okamura2004}. We observe a similar shoulder in $\sigma_{aa}$ at $50\,$meV, which emerges at $100\,$K – close to the coherence temperature of $70\,$K obtained from $\rho_{a}$ \cite{kim2003}. It is interesting that this feature is not present in $\sigma_{cc}$ where DC resistivity data is similarly found to lack a coherence peak for $\rho_{c}$. 

The fact that a coherence peak is observed in $\rho_{a}$ but not $\rho_{c}$ suggests that the quasiparticles which dominate the $c$-axis transport have a significantly smaller hybridization to the $f$-electrons than the quasiparticles contributing to the in-plane transport. This anisotropy is reflected in our extended Drude analysis. While the effective mass $m^\ast$ and the frequency dependent scattering rate $\Delta\Gamma$ are still enhanced in the $c$-axis conductivity data, the enhancement is at least twice as strong for the $a$-axis conductivity. Remarkably, performing the same extended Drude analysis on the theoretically computed optical conductivity reproduces the frequency and temperature dependence for both directions down to $10\,$meV (see Figs.~\ref{extDrudeParameters}(g-j)). This suggests that the anisotropic mass renormalization is fundamentally a property of the underlying band structure as well, which is renormalized through a momentum independent self energy of the $f$-electrons.

We have illustrated that the electronic structure and the renormalization thereof in CeRhSn is strongly anisotropic and spans energy scales up to $0.8\,$eV, confirming CeRhSn as a mixed valent material. Despite the presence of magnetic frustration in the Ce-lattice, the anisotropy of the renormalization and the energy scales can be well accounted for from first-principles calculations using a DFT+DMFT framework, which assumes a momentum independent self energy. Several mysteries in CeRhSn remain. Notably, it is highly counterintuitive to observe a strongly anisotropic spin susceptibility in a mixed valent compound for which no high energy crystal field excitations have been identified \cite{higaki2006}. Also surprising is the observation of the non-Fermi liquid behavior at low temperatures, which appears to be related to the presence of magnetic frustration \cite{tokiwa2015}. Additional spectroscopic measurements to lower frequencies would be interesting to clarify the origins of these puzzles. 

We thank Ken Burch, Peter Armitage, Ken O'Neal and Rohit Prasankumar for enlightening discussions to improve our FTIR spectroscopy effort. Work at Los Alamos was carried out under the auspices of the U.S. Department of Energy (DOE) National Nuclear Security Administration (NNSA) under Contract No. 89233218CNA000001. The experimental work acknowledges support from the DOE Office of Basic Energy Sciences, Materials Sciences and Engineering Division. The theoretical calculations were performed with support from LANL LDRD Program with project XXYU and XXNV. This research used the Infrared Lab of the National Synchrotron Light Source II, a U.S. DOE Office of Science User Facility operated for the DOE Office of Science by Brookhaven National Laboratory under Contract No. DE-SC0012704. 

{\it Note added}: While completing this manuscript we became aware of a related work \cite{kimura2024}. The optical conductivity data is in very good agreement with ours though the focus of their manuscript is different.

\bibliographystyle{apsrev4-1}
\bibliography{refs}

\clearpage
\newpage
\pagebreak
\widetext
\begin{center}
\textbf{\large Supplementary material for ``Anisotropic hybridization in CeRhSn"}
\end{center}

\setcounter{equation}{0}
\setcounter{figure}{0}
\setcounter{table}{0}
\setcounter{page}{1}
\makeatletter
\renewcommand{\theequation}{S\arabic{equation}}
\renewcommand{\thefigure}{S\arabic{figure}}
\renewcommand{\bibnumfmt}[1]{[S#1]}
The sample surface was prepared by mechanically polishing parallel to the $ac$-plane to a roughness of ${\sim}0.3\mu\mathrm{m}$. The correct orientation of the applied electric field with respect to the crystallographic axes was verified by the phonon spectrum in Fig.~\ref{supplement1}(a). The optical phonons in $\sigma_{aa}$ and $\sigma_{cc}$ are clearly visible and at distinct energies. The Laue pattern of Fig.~\ref{supplement1}(b) shows that the polished surface of the sample is an $ac$-plane. Laue patterns were obtained at 14 positions on that $ac$-plane, as depicted in Fig.~\ref{supplement1}(c). Except for one position at the lower left edge, the uniformly oriented Laue results demonstrate a single grain throughout the plane. Clear peaks indicate a low degree of disorder. 

\begin{figure}[ht]
\includegraphics{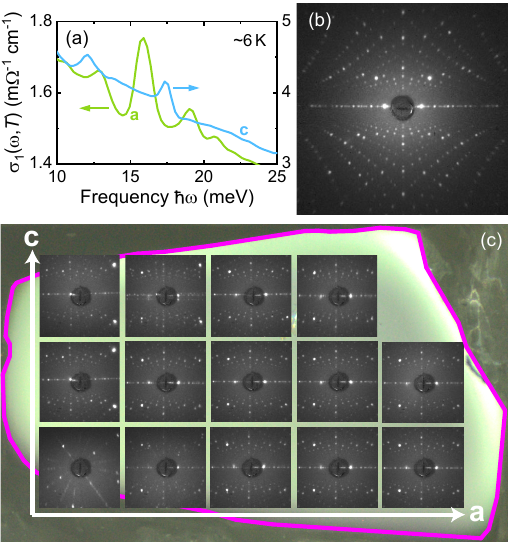}
\caption{Sample orientation and homogeneity. Panel (a) shows the phonon spectrum of $\sigma_{aa}$ (green) and $\sigma_{cc}$ (blue), panel (b) depicts a Laue image after the final polish from the center of the sample, and panel (c) compiles 14 Laue scans on different positions on the sample. The $ac$-surface of the sample is the shiny green area encircled with a pink line for illustrative purpose.}     
\label{supplement1}
\end{figure}

\begin{figure}[ht]
\includegraphics{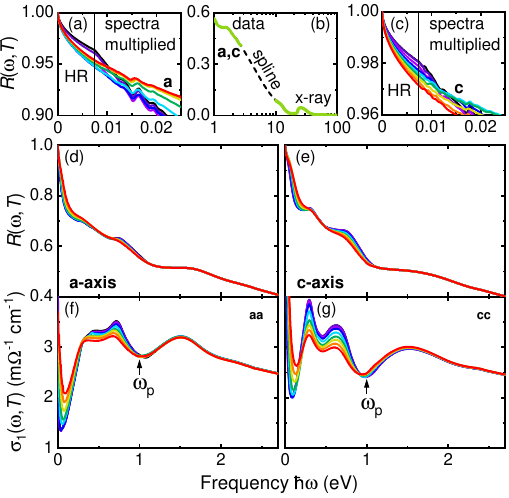}
\caption{Extensions and results for the full measured frequency range. Low energy extensions are shown in (a) and (c) for the $a$ and $c$ axis, respectively. The vertical line at $7.4\,$meV separates the range where the Hagen-Rubens formula (HR) is used and where the measured and multiplied spectra are taken. The high energy extension can be seen in (b) for both $a$ and $c$. Reflectivity spectra are plotted in (d) and (e), the corresponding optical conductivity in (f) and (g). The arrows mark the plasma frequency at $1\,$eV.}     
\label{supplement2}
\end{figure}

\begin{figure}[ht]
\includegraphics{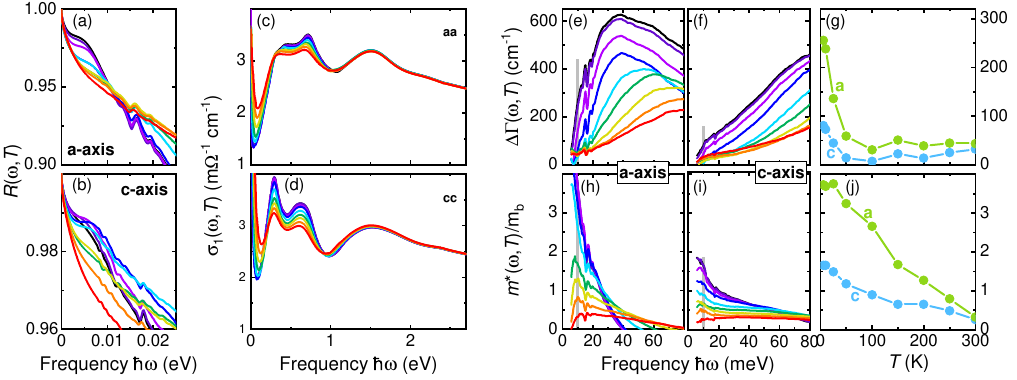}
\caption{Results for another method of extending the spectra to low energies. Low energy extensions of the reflectivity are shown in (a) and (b) The Hagen-Rubens formula is applied for $\omega \rightarrow 0$ which turns into a linear function that continues the measured spectra to energies below $7.4\,$meV. This yields the optical conductivity shown in (c) and (d) and the scattering rates and effective masses as in (e-g) and (h-j), respectively.}     
\label{supplement4}
\end{figure}

Reflectivity spectra as compiled in Fig.~\ref{supplement2} were recorded at temperatures from ${\sim}6\,$K to $300\,$K in a frequency range of $6\,$meV to $1.0\,$eV and at $300\,$K from $1.0\,$eV to $2.7\,$eV, where a lack of temperature dependence was evident for these higher energy interband transitions. We employed an energy resolution of $\Delta E = 0.5\,$meV from $6\,$ to $60\,$meV, $\Delta E = 1.2\,$meV up to $1.0\,$eV and $\Delta E = 2.5\,$meV for $1.0\,$eV and above. All data were collected on a Bruker Vertex 80v FTIR spectrometer, where absolute reflectivity was determined following \textit{in situ} gold ($<1.0\,$eV) or silver ($>1.0\,$eV) evaporation to account for the ${\sim}0.3\mu\mathrm{m}$ surface roughness resulting from polishing \cite{homes1993}. In order to obtain the optical conductivity from the reflectivity, the applied Kramers-Kronig transform requires both a low ($<\,6\,$meV) and high ($>\,2.7\,$eV) frequency extension into regions which are inaccessible in our experiment. At high frequencies, reflectivity spectra were extended up to $30\,$keV by a cubic spline, connecting our data with the x-ray reflectivity computed by model calculations \cite{tanner2015, henke1993, gulliksonWebsite}. Low frequency ($< 10\,$meV) extensions relying on a conventional Hagen-Rubens by $R(\omega,T)=1-2\sqrt{\epsilon_0 \omega \rho_\mathrm{dc}(T)}$ as shown in Fig.~\ref{supplement2} and phenomenological (linear extrapolation to ${\sim}5\,$meV, which is then merged with the Hagen-Rubens extension) as in Fig.~\ref{supplement4} were applied using the experimental, temperature dependent DC resistivities provided in Ref.~\cite{kim2003}. While neither model captures the behavior of our reflectivity spectra below $10\,$meV (error $< 2$\%), we note our extended Drude analysis to be robust against this choice of extrapolation, leading to an overall confidence in our conclusion for the scattering rate and effective mass anisotropy at ${\sim}6\,$K to fall within an error of $< 3$\% and $< 8$\%, respectively.

\begin{figure}[ht]
\includegraphics{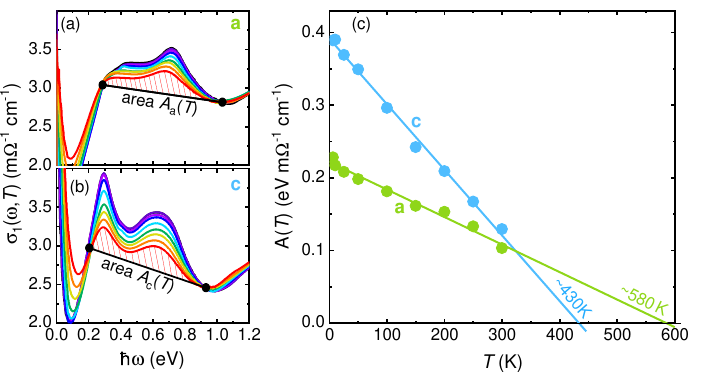}
\caption{Extrapolation of the area of the hybridization-induced peaks. The area of the peaks was estimated as illustrated in (a) and (b). A linear model background (black line) connects the approximate isosbestic points (black dots) of the optical conductivity, which was subtracted from the spectra to obtain the peak areas. As an example, the area for $300\,$K is hatched in red. The resulting areas are plotted as a function of temperature in (c), together with linear extrapolations for the $a$ and $c$-axis in light green and blue, respectively.}     
\label{supplement3}
\end{figure}

To estimate the temperature scales of the hybridization, we extrapolated the hybridization-induced peak areas observed in the optical conductivities to temperatures above $300\,$K as described in Fig.~\ref{supplement3}. This yields ${\sim}430\,$K and ${\sim}580\,$K for the hybridization-induced peaks to vanish completely in the $c$ and $a$-axis, respectively. A decay of the area to half of the value at the lowest temperature is reached in the range $200-300\,$K, in agreement with reported values \cite{kim2003}.

The performed electronic structure calculations use fully charge self-consistent DMFT calculations combined with DFT and are implemented in the eDMFT functional code~\cite{haule2010}. DFT calculations were performed using the WIEN2k code, which uses a full potential augmented plane-wave method \cite{Blaha2020}. A 2000 $k$-point mesh was used for self-consistent calculation. A hybridization window was set from $-10\,$eV to $10\,$eV with respect to the Fermi level ($E_{F}$). The rotational invariant Slater form of Coulomb interaction was used in the calculation. The nominal double counting method was applied, where the nominal occupancy of the Ce atom was set to 1.

\begin{figure}[ht]
\includegraphics{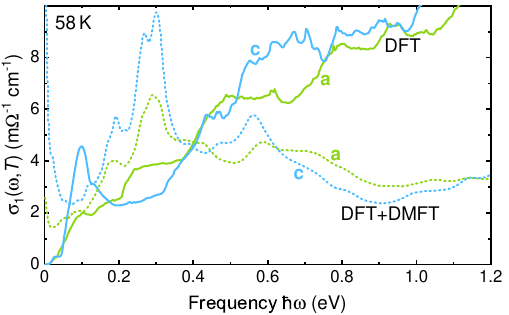}
\caption{Optical conductivity from DFT calculation. For comparison, the optical conductivity from DFT+DMFT calculations as in Figs.~\ref{theory}(c) and (d) are plotted as dotted lines. The Drude part is not included in DFT optical conductivity.}     
\label{supplement5}
\end{figure}

Fig.~\ref{supplement5} shows a comparison of the optical conductivity from DFT+DMFT (as presented in Fig.~\ref{theory}) with a calculation from DFT alone. While the anisotropy in the electronic structure is reflected by the DFT, the importance for including DMFT, which accounts for local correlations, is apparent.  

\end{document}